\documentclass{article}

\usepackage{todonotes}
\usepackage{lineno,hyperref}
\modulolinenumbers[5]
\usepackage[margin=1in]{geometry}

\title{Hard Choices in Artificial Intelligence}

\author{Roel Dobbe\thanks{Faculty of Technology, Policy and Management, Delft University of Technology. \texttt{r.i.j.dobbe@tudelft.nl}} , Thomas Krendl Gilbert\thanks{Center for Human-Compatible AI, UC Berkeley. \texttt{tg340@berkeley.edu}}, Yonatan Mintz\thanks{Department of Industrial and Systems Engineering, University of Wisconsin-Madison. \texttt{ymintz@wisc.edu}} \footnote{All authors contributed equally to this work.}}

\begin{document}
\maketitle

\begin{abstract}
As AI systems are integrated into high stakes social domains, researchers now examine how to design and operate them in a safe and ethical manner. However, the criteria for identifying and diagnosing safety risks in complex social contexts remain unclear and contested. In this paper, we examine the \textit{vagueness} in debates about the safety and ethical behavior of AI systems. We show how this vagueness cannot be resolved through mathematical formalism alone, instead requiring deliberation about the politics of development as well as the context of deployment. Drawing from a new sociotechnical lexicon, we redefine vagueness in terms of distinct design challenges at key stages in AI system development. The resulting framework of \textit{Hard Choices in Artificial Intelligence} (HCAI) empowers developers by 1) identifying points of overlap between design decisions and major sociotechnical challenges; 2) motivating the creation of stakeholder feedback channels so that safety issues can be exhaustively addressed. As such, HCAI contributes to a timely debate about the status of AI development in democratic societies, arguing that deliberation should be the goal of AI Safety, not just the procedure by which it is ensured. \\
\textit{Keywords:}
AI Governance, AI Safety, Ethics \& Philosophy of Artificial Intelligence, Sociotechnical Systems, Cybernetics
\end{abstract}

\section{Introduction}

The rapid adoption of AI systems is reshaping many public, professional, and personal domains, providing opportunities for innovation while also generating new forms of harm.
These harms are diverse, ranging from physical dangers related to new robotic systems (e.g. autonomous vehicles~\cite{hawkins_serious_2019}), to economic losses related to welfare systems~\cite{henley_welfare_2020}, to forms of racism and discrimination in systems that engage with biometrical data in public spaces~\cite{hill_wrongfully_2020,harmon_as_2019} or with personal data on social media platforms~\cite{hao_facebooks_2019,hao_he_2021}.
These cases reveal emerging gaps between the promised beneficial outcomes of AI applications and the actual consequences of deployed systems. 
Ongoing risks and harms are thus a product of the \emph{sociotechnical gap}, ``the great divide between what we know we must support socially and what we can support technically''~\cite{ackerman2000intellectual}.

In response, a broad spectrum of civil society initiatives have emerged to safeguard human domains from the effects of AI systems.
Debates about the sociotechnical gap have taken two forms.
One is the proposal of normative principles to determine how the gap should be filled or who should do it. 
This has led to a plethora of reports and statements~\cite{schiff_ai_2021} about how AI should be governed to respect fundamental rights~\cite{andersen_human_2018,noauthor_getting_2020}, alongside a growing need to operationalize these principles~\cite{mittelstadt_principles_2019}. 
For example, the OECD Principles on Artificial Intelligence ``promote artificial intelligence (AI) that is innovative and trustworthy and that respects human rights and democratic values,'' and are signed by governments~\cite{noauthor_recommendation_2021}.
The European Commission recently proposed a regulatory framework to translate higher-level principles into concrete technical and legal solutions through ``harmonized standards''~\cite{noauthor_proposal_2021}.
However, it is unclear how these standards could reconcile the diverse needs of users in the context of particular systems and domains.
Second is the proposal of technical tools to better fill the gap. 
While these efforts have generated many technical approaches related to mathematical criteria for ``safety'' or ``fairness'', their systematic organization and prioritization remains unclear and contested ~\cite{gebru_datasheets_2020,mitchell_model_2019,raji_actionable_2019,green_algorithmic_2020}.

Missing from both debates is a sustained interrogation of \textit{what it means} to identify, diagnose, and ultimately fill the sociotechnical gaps generated by AI systems. This entails asking deeper questions about how a given system may restructure human values and social practices, whether technical and governance criteria may be reconciled in design choices, and when or where gaps emerge across the system's development lifecycle. Put differently, we lack a presentation of AI development as a deliberative practice that interrogates how what present and future AI systems will be capable of bears on what it will mean for these systems to work for us.


Concretely, every AI system requires a consensus definition of what it would mean for it to be safe. 
But present proposals for the technical safety and governance of AI systems tend to focus on safety either as a criterion of technical design, operational conditions, or the experience of end users.
This means safety criteria are marred by an underlying \emph{vagueness}, the absence of unifying categories to establish whether a system’s capabilities are safe or not. 


This paper makes two key claims. First, AI development must be reconceived in terms of the multiple points of encounter between system capabilities and sociotechnical gaps. This requires a new vocabulary and framework to make sense of salient gaps in the context of technical design decisions, constituting a reciprocal relationship between system development and governance. Second, developers must take on new roles that are sensitive to feedback about how to manage these gaps. This requires communicative channels so that stakeholders are empowered to help shape the criteria for design decisions.

Our contributions flow from these two claims. In Section 2 we supply a lexicon of terms for the problems at stake in sociotechnical gaps. In Section 3 we analyze the present landscape of proposed technical and normative solutions to particular gaps in terms of piecemeal responses to vagueness. In Section 4 we present Hard Choices in Artificial Intelligence (HCAI) as a systematic framework that maps possible gaps to particular feedback channels for designers and stakeholders to use. In Section 5 we present this framework’s implications for designers and advocates when evaluating the technical performance and governance standards of actual systems. Section 6 concludes.

We emphasize that our concerns, while responding to more recent iterations of AI and computer systems, are not new. The research agenda of situated design \cite{greenbaum_design_1992} and Agre’s call for a ``critical technical practice'' \cite{agre_computation_1997} comprise classic phenomenological critiques of ``good old-fashioned'' symbolic and expert systems, in particular the need to become critical about certain formal assumptions behind intelligence and to reassess problematic metaphors for perception and action \cite{dreyfus2014skillful}. Yet much technical research today has moved beyond these critiques. Reinforcement learning (RL), for example, reflects Dreyfus’ exposition of intelligence as a learned, situated, dynamic activity developed from coping with one’s surrounding environment and embodying different strategies for action. 
The question is no longer what computers can or cannot do, but \textbf{how to structure computation in ways that support human values and concerns}. To support this aim, we propose AI practitioners will need new \textit{cybernetic practices} that guide how feedback may be solicited from existing and emerging political orders. 

We thus apply an insight to AI development that scholars in Science and Technology Studies (STS) have appreciated for over four decades: any and every technological system is political, requiring normative deliberation and ongoing collective participation to ensure its safety for everyone affected by it \cite{winner1980artifacts}.

\section{Towards a Sociotechnical Lexicon for AI}
\label{sec:lexicon}
At present, AI research lacks a robust sociotechnical lexicon. This would include the emerging problem space of AI Safety as well as newly-relevant questions of 
cybernetics in the context of present and future AI governance topics. In this section we present a preliminary lexicon to reveal areas of overlap and divergence between these domains, enabling comparison between contemporary assumptions of AI development and possible alternative paradigms.

As was stated in the original Dartmouth summer project proposal, research on artificial intelligence is meant to pursue ``the conjecture that every aspect of learning or any other feature of intelligence can in principle be so precisely described that a machine can be made to simulate it'' \cite{mccarthy2006proposal}. Beneath specific efforts to simulate language, brain models, and intellectual creativity, AI theorists were most interested in precision: adequately specifying the mechanisms underpinning intelligence such that they would be possible to replicate via computation and symbolic reasoning. This quest for exactness has continued to underpin many technical and conceptual interventions on how to model the intelligent behavior of agents within some environment, including the problem of specification in reinforcement learning \cite{milli2017should,hadfield-menell_inverse_2017}.

\begin{itemize}
    \item agency--the capacity of some agent (human or artificial) to act in order to achieve a particular outcome or result.
    \item intelligent agent (IA)--an autonomous entity which acts, directing its activity towards achieving goals.
    \item environment--a domain in which an IA can perceive through sensors and act using actuators, in pursuit of a goal.
    \item AI model--a mathematical representation of the environment, constructed through either simple rules, a model, or a combination thereof, the parameters of which may be learned from and updated with observed data.
    \item objective function--a mathematical representation capturing the goals of the IA.
    \item specification--the definitions of the environment, the IA's sensors and actuators, and the internal model and objective function necessary to operate and (learn to) perform a particular task.
    \item artificial intelligence--the study of how to design IAs that simulate, approximate, or surpass the precise capabilities of human intelligence.
\end{itemize}

In recent years, the rapid advent of AI functionality across societal domains has motivated the formulation of principles and definitions that consider such artifacts in their system setting. Here we include definitions adopted by the OECD in 2019~\cite{noauthor_recommendation_2021}.

\begin{itemize}
    \item AI system--a machine-based system that can, for a given set of human-defined objectives, make predictions, recommendations, or decisions influencing real or virtual environments. AI systems are designed to operate with varying levels of autonomy.
    \item AI system lifecycle--involves: i) ‘design, data and models’; which is a context-dependent sequence encompassing planning and design, data collection and processing, as well as model building; ii) ‘verification and validation’; iii) ‘deployment’; and iv) ‘operation and monitoring’. These phases often take place in an iterative manner and are not necessarily sequential. The decision to retire an AI system from operation may occur at any point during the operation and monitoring phase.
    \item AI knowledge--the skills and resources, such as data, code, algorithms, models, research, know-how, training programmes, governance, processes and best practices, required to understand and participate in the AI system lifecycle.
    \item AI actors-those who play an active role in the AI system lifecycle, including organisations and individuals that deploy or operate AI.
    \item stakeholders--all organisations and individuals involved in, or affected by, AI systems, directly or indirectly. AI actors are a subset of stakeholders.
    \item stakeholder--a person or entity with a vested interest in the AI system's performance and operation.
\end{itemize}

Today, this system lens to AI is largely inspired by the field of ``AI Safety'' and the associated technical project of ``value alignment'', which aims to build ``provably beneficial'' systems that learn the precise preference structures of humans \cite{russell2017provably}. Value alignment assumes that such a deterministic description already exists or is discoverable by artificial agents, and if we create precise mechanisms for learning it, then it could be modeled under mathematical conditions of uncertainty.

\begin{itemize}
    \item AI Safety--the interdisciplinary study of how to build systems that are aligned with the structure of human values, in particular those of stakeholders whom the system is meant to serve.
    \item value alignment--the creation of systems whose specification is sufficient to learn the structure of human values.
\end{itemize}

%
In practice, AI research is as much about redefining philosophical concepts in the context of AI as it is about solving particular engineering and computer science challenges. But there is a fundamental gap between the idea of value alignment and managing the actual consequences of deployed systems. 
Decades of research in systems engineering for safety-critical systems has shown that values, such as safety or fairness, are an \emph{emergent} property that ``arise from the interactions among the system components''~\cite{leveson_engineering_2012}. Here, the system boundary and its components entail both technical elements or intelligent agents, as well as human agents, processes and supporting infrastructure. 
\begin{quote}
    ``The emergent properties are controlled by imposing constraints on the behavior of and interactions among the components. Safety then becomes a control problem where the goal of the control is to enforce the safety constraints. Accidents result from inadequate control or enforcement of safety-related constraints on the development, design, and operation of the system.''~\cite{leveson_engineering_2012}
\end{quote}
The emergent and dynamic nature of values, and the inability to ``discover'' or formalize these in the technical logics of a system, is corroborated by a long tradition of research in computer-supported cooperative work~\cite{greenbaum_design_1992}, human-computer interaction~\cite{shilton_values_2018} and participatory design~\cite{halloran_value_2009}. As Halloran et al. conclude, ``values emerge, whether you look for them or not''~\cite{halloran_value_2009}.
This problem resonates with the classic problem space of cybernetics: continuously interrogating and elaborating the relationship between actions and goals through forms of feedback rather than a deterministic problem formulation or static representation of value \cite{wiener1988human,von2007understanding}.
%
%
In cybernetics, performance thresholds are determined through the concrete outcomes of actions taken, rather than precisely-defined capacities of the agent that reflect some stylized view of intelligence. This entails looking at the level of systems, composed of integrated components, and how values such as safety are instantiated and maintained through conditions of stability.

\begin{itemize}
    \item feedback--information about the results of an agent’s or system’s actions which can then be taken as inputs for future actions, serving as a basis for improvement or stability.
\item cybernetics--the interdisciplinary study of how systems behave in response to feedback.
\end{itemize}
As pragmatist philosophers and sociotechnical scholars have long emphasized, bridging the gap between design principles and real-world system performance requires specifying the normativity of the problem domain in terms of acceptable behaviors and outcomes \cite{pask1976conversation,dewey1896reflex}. On this interpretation, cybernetic feedback is needed to bridge the gap between problem formulation and defining the system's interface with reality. The question is: are norms something that can be passively learned by an agent or something enacted through new forms of feedback? The former implies uncertainty about norms that in principle could be modeled by e.g. learning a reward function that represents human preferences. The latter however suggests indeterminacy that cannot be resolved without a broader system lens to instantiate design or governance norms.

Recent work in AI Governance suggests the latter. As argued by Wallach and Marchant, the most pressing regulatory questions will require new institutional entities tasked with articulating metrics, standards, or new forms of domain expertise to determine acceptable performance thresholds for particular AI systems \cite{wallach2019toward}. This may include governance coordination committees \cite{cihon2019standards}, an International Artificial Intelligence Organization \cite{erdelyi2018regulating}, the Facebook Oversight Board \cite{klonick2019facebook}, judicial oversight in the spirit of the EU General Data Protection Regulation \cite{voigt2017eu}, issues studied by the National Institute of Standards and Technology \cite{smuha2021race}, ``arms race'' scenario modeling \cite{zwetsloot2018beyond}, and many others. This emerging literature seeks to resolve situations of indeterminate system performance at various levels of normative abstraction, ranging from individual privacy to global security concerns.

\begin{itemize}
    \item sociotechnics--the relationship between a system and real-world conditions, whose specification requires active engagement with the concerns of stakeholders.
\item normativity--the reciprocal expectations of agents to conform to particular agreed-upon standards of behavior in a given domain.
\item normative uncertainty--the unknown features of an environment that the agent must learn in order to behave optimally.
\item normative indeterminacy--the lack of prior standards or forms of consensus for the sociotechnical context of a given system, rendering the specification problematic or incomplete.
\end{itemize}

Contemporary sociotechnical concerns about the development of AI systems share a common theme: data accumulation, increasing computational capacity, and new algorithmic learning procedures are reconstituting the normative systems in which humans live \cite{yeung2017hypernudge,seaver2019knowing,gillespie2014relevance}. In this sense, the problem space of AI Safety is rediscovering cybernetics on new ground. There is an emerging need for sociotechnical specifications that are able to diagnose and resolve undesirable system performance, semantic equivocations, and political conflicts. This requires a principled elaboration through which an AI system's technical specifications (i.e. its model, objective function, sensors, actuators) are interpreted in light of salient normative considerations and real-world performance thresholds that stem from the social and situated context in which the system operates. Without clarifying this landscape, it will not be possible to evaluate whether particular governance mechanisms at different institutional scales are more or less appropriate for addressing the indeterminacies at stake.

\begin{itemize}
    \item featurization--the system’s capacity to represent features of the environment in order to achieve a specified goal.
    \item optimization--the designer’s capacity to articulate how to more efficiently (e.g. cost minimization) or appropriately complete a task.
    \item integration--the capacity of users, managers, regulators and stakeholders to oversee and incorporate the system’s real-world performance.
    \item sociotechnical specification--the proposed normativity of an AI system in terms of its featurization, optimization, and integration, defining who it is meant to serve, its purpose, and how it is to be evaluated and held accountable.
\end{itemize}

As Erdelyi and Goldsmith note, ``the choice between harder and softer types of legalization [of AI systems] involves a context-dependent tradeoff, which actors should carefully consider on a case-by-case basis'' \cite{erdelyi2018regulating}. To weigh such tradeoffs, it must first be possible to index values and norms in terms of technical decisions about the system specification. This means that normative concerns of comparable significance and scope must be rendered commensurable in order for a responsible tradeoff to be struck and translated to a system's specification. 
Ruth Chang has highlighted the related philosophical notion of parity \cite{chang1997incommensurability, chang_possibility_2002}, which holds that humans are able to articulate evaluative differences to make comparisons between incommensurable values or options \cite{chang2017hard}. This permits practical deliberation regarding one's overarching goals. Parity is constitutive of what Chang calls \textit{hard choices}: when different alternatives are on a par, ``it may matter very much which you choose, but one alternative isn't better than the other [...] alternatives are in the same neighborhood of value [in terms of how much we care] while at the same time being very different in kind of value''.
Note that while Chang developed the notion of parity and hard choices for an individual agent or authority weighing different options or values, we reinterpret these concepts in a setting comprising different stakeholders. 
This renders the weighing of options or values, and thereby notions of parity and hard choices, as inherently \emph{political} as different stakeholders will have different interests, varying political power and potentially diverging ideas about evaluating different problem formulations, solution directions and associated values or principles~\cite{de_haan_solving_2015}.
We acknowledge recent empirical insights from Van der Voort et al., who debunk the rational view typically assumed for for decision-making. They show how algorithms and big data analytics encounter political and managerial institutions in practice, leading to a spectrum of possible outcomes or theses for how the technology is specified and used~\cite{van_der_voort_rationality_2019}. 

\begin{itemize}
    \item comparability--the evaluation of an AI system's technical capacities (e.g. learnable features) as similar to each other in their magnitude, relevance, or problem stakes.
    \item incommensurability--the evaluation of an AI system's normative capacities (e.g. relationship with users or designers) as not able to be measured by the same standard.
    \item parity--A relation between values that are comparable in significance but unable to be directly measured as better, worse, or equal to each other.
    \item hard choices--Situations of value parity in AI system development, which require normative deliberation in order to make the options technically commensurable.
    \item politics of hard choices--activities and behaviors associated with the stakes involved in hard choices, especially the debate and interactions between parties having or seeking power over its resolution.
\end{itemize}

The possibility of hard choices when designing AI systems suggests the need for a principled diagnostic approach, folded into development practices. This approach would specify commitments that match appropriate modalities of algorithmic governance with the potential harms faced by stakeholders. The goal would not be for developers to make choices on stakeholders' behalf, but for developers to adopt diagnostic practices so that choices can be proactively anticipated and resolved through feedback. As argued by Elizabeth Anderson \cite{anderson2006epistemology}, the form of feedback particular to modern democracies is dissent, indicating that the current specification (e.g. of a law) is problematic and must be amended or rejected. Accommodating dissent is thus a path to enacting appropriate features as well as proportional mechanisms for democratic governance, denoting a possible alternative form of design practice.

\begin{itemize}
\item commitment--a pledge made by developers to stakeholders about the sociotechnical specification of an AI system, in terms of how it is intended to operate.
\item dissent--purposive feedback that lies outside the distribution of previous inputs, serving to challenge the grounds for consensus on system specification.
\item cybernetic practice--active attention to the types of feedback needed to address normative indeterminacies and refine the sociotechnical specification of a particular AI system.
\end{itemize}

Revealingly, 
the field of cybernetics also applied feedback to cybernetic practices themselves, which culminated in so-called "second-order cybernetics"~\cite{glanville_purpose_2004}. 
We embrace the spirit of this tradition, as well as later work proposing such reflective inquiry on technical practices in AI~\cite{agre_toward_1997}.

From this lexicon, we conclude that recent work in AI governance and AI Safety reveals a need for:

\begin{enumerate}
    \item a sociotechnical reframing of classic problem domains in AI (agency, models, representation, learning), in terms of how human behaviors and institutions will be indeterminately reshaped by designed systems.
    \item a shared language to diagnose different kinds of normative indeterminacy, both between intended vs. actual system behavior and across communities of stakeholders.
    \item the specification of requisite feedback modalities, in order for the system to achieve appropriate stability in the face of operational indeterminacies.
\end{enumerate}

\section{The Problem of Vagueness}
\label{sec:vagueness}
%
As AI systems are applied to more sensitive contexts and safety-critical infrastructure, normative indeterminacies are becoming more visible. Identifying the missing feedback in a given specification requires interrogating the functions of an AI system in a principled manner. This includes examining what task the AI system is trying to complete and how the system is meant to work in support of human contexts, as well as which normative standards would be appropriate to fulfill these needs. 

Here we compare prominent technical and policy standards that have been proposed, revealing each as a partial response to the underlying problem of \textit{vagueness}. 
The vagueness of a system specification is the ultimate source of the normative indeterminacies at stake. Vagueness is a central topic in metaphysics and the philosophy of logic and language that has important application in system engineering and artificial intelligence \cite{agre_toward_1997}. 
It is about the fundamental lack of clarity in our relationship with the world, either in terms of the ways we are able to perceive it, the language we use to describe it, or in the world itself.
It is addressed through the drawing of boundaries--forms of classification, demonstration, analogy, and other rhetorical strategies that sort phenomena into particular qualities and quantities or draw distinctions of form and content \cite{williamson2002vagueness}. A classic example is the Sorites paradox: which grain of sand removed from a heap turns the heap into a non-heap? Such situations may yield existential uncertainty, which, if not resolvable through agreed upon standards, may lead to arbitrary tradeoffs, compromise, or restrictions. We thus propose vagueness as a general descriptor for situations in which developers’ attempts to model some domain via technical uncertainty fall short and give way to specific forms of indeterminacy.

For each approach to indeterminacy present in the current AI policy and governance literature, we first organize and present the corresponding classical interpretation of vagueness, namely either \textit{epistemicism}, \textit{ontic incomparabilism}, or \textit{semantic indeterminacy}~\cite{chang_possibility_2002}. We then isolate the respective standards that have been unreflectively derived from these schools of thought, namely \textit{metanormativism}, \textit{value pluralism}, and \textit{fuzziness}. Finally, we identify the stylized form of feedback that each school of thought prioritizes over others to enact these standards, namely \textit{preference learning}, \textit{refusal}, and \textit{equitable outcomes}. We concisely summarize these relationships in Table 1. This exercise motivates the need for sustained engagement with the actual context of system development. 

\subsection{Epistemicism - resolving vagueness through model uncertainty}

\textit{Epistemicism} claims bivalence as a basic condition for an object’s existence \cite{schiffer1999epistemic}. This is to say that for any given property of an object, there is in principle some sharp boundary by which the object either does or does not have that property. 
Illustrated through the Sorites paradox, epistemicists believe that there is an objective fact of the matter about the precise number of sand grains necessary to constitute a heap vs. non-heap, even though we may be ignorant of that cutoff point. 
The position thus holds that every object property or attribute must terminate at some boundary, no matter how inappreciable this boundary may be at present. 
This implies that acquiring more information may help reveal where the boundary actually is or could be drawn. 
Pure epistemicism is counterintuitive and is philosophically controversial in comparison with the claim that boundaries are semantic constructions \cite{gomez1997two}. But the essence of the position is simply that if distinct communities (or even the same person) claim the same property applies to the same object in different ways, then they are either ignorant about the property’s actual boundary or are describing distinct objects.

Epistemicism has a powerful affinity with \textit{metanormativism}, the notion that the criteria for rational decision-making are not fully known or confidently expressed because sufficient information about the environment, other agents, or oneself is absent. 
Because epistemicists believe that no comparable options are fundamentally ``apples and oranges'', as there must be some degree to which one is preferable over the other, metanormativism asserts the existence of a clear, positive value relation between available ethical actions: one must be unambiguously better, worse, or equal to the other for a given choice to be demonstrably rational. 
For example, William MacAskill has sought to articulate ``second-order norms'' that guide how one should act when multiple appealing moral doctrines are available \cite{macaskill2019practical}. 
MacAskill, whose work has been cited in support of technical work on AI value alignment and value learning \cite{soares2014aligning, soares2015value}, has also proposed a ``choice worthiness function'' that would generate reward functions in an ``appropriate'' manner, where appropriateness is defined as ``the degree to which the decision-maker ought to choose that option, in the sense of ‘ought’ that is relevant to decision-making under normative uncertainty'' \cite{macaskill2016normative}. 
As such, metanormativism is a natural ally of expected utility theory and in particular the first axiom of the Von Neumann-Morgenstern utility theorem, specifying the completeness of an agent’s well-defined preferences \cite{von2007theory}.

Distinct approaches to AI Safety have emerged to define the uncertain scale at which AI systems may cause social harm. 
At one end of this continuum is \emph{existential risk} (hereafter referred to as x-risk), i.e. the effort to mathematically formalize control strategies that help avoid the creation of systems whose deployment would result in irreparable harm to human civilization. 
The x-risk literature has focused on the ``value alignment problem'' in order to ensure that learned reward functions correspond with the values of relevant stakeholders (such as designers, users, or others affected by the agent's actions) \cite{soares2015value}. Here the reward function serves as a representation of stakeholder preferences rather than the AI agent's own objective function, an assumption common in inverse reinforcement learning \cite{hadfield-menell_cooperative_2016}. This position is also practically adopted by software engineers and tech enthusiasts for whom the uncertain specification of human preferences comprise an investment opportunity for new AI systems. 
The following quote from Mark Zuckerberg is illustrative: ``I’m also curious about whether there is a fundamental mathematical law underlying human social relationships that governs the balance of who and what we all care about [...] I bet there is'' \cite{hildebrandt2019privacy}. 

The promise of such a function continues to provide guidance for designers and AI researchers about what decision procedures are acceptable or unacceptable for the system to follow, specifically when the goal state and risk scale are difficult to define \cite{hadfield2019incomplete, irving2019ai}. 
This research agenda prioritizes \textit{preference learning}, the systematic observation of user behavior and choices to learn an underlying reward function, as the most salient form of design feedback for filling the gaps in system specification \cite{hadfield-menell_cooperative_2016}. 
As stated by Stuart Russell \cite{russell2019human}: 
\begin{itemize}
    \item The machine's only objective is to maximize the realization of human preferences.
    \item The machine is initially uncertain about what those preferences are.
    \item The ultimate source of information about human preferences is human behavior.
\end{itemize}
However, this vision is inadequate for design situations in which human behavior is difficult to observe. Reasons for this could be empirical (sparse behavioral signals) or normative (concerns about surveillance or behavioral manipulation).

\subsection{Ontic incomparabilism - respecting value pluralism}

Meanwhile, \textit{ontic incomparabilism} holds that there are fundamental limits to what our predicates or semantics can make of the world because there is no objective basis to prefer one definition of a concept to another \cite{barnes2011theory}. 
More concretely, even if we knew everything about the universe, there would still be no way to argue that a pile of sand ``should be considered a heap'' after exactly n+1 grains as opposed to after n grains. Ontic incomparabilism therefore claims that we cannot ever fully model the world by discovering additional criteria or accumulating sufficient information about it as its dynamics may be fundamentally unsuited to model specification. Note that this position is distinct from views that the world is impossible for human minds to comprehend completely (as has been argued for specific physical phenomena, e.g. quantum mechanics) or that the world is impossible to describe accurately. Instead, the claim is that any finite number of descriptions or representations cannot exhaust the world’s richness because its basic features are not readily discernible, and that there are in principle as many different ways of representing the world as there are agents capable of realizing their agency in that world. This means that modeling the world robustly would require securing the world’s total cooperation with the boundaries being drawn over it.

Ontic incomparabilism has found expression in terms of \textit{value pluralism}, i.e. that there cannot or will never be an ultimate scheme for delineating human values because humans exist in the world in a way that cannot be exhaustively represented. This transcends sociological fact (i.e. that people hold different beliefs about values, and value beliefs differently) to make an axiological, anti-monist claim: values are indeterminately varied and incommensurable, and no ethical scheme could ever account for the range of values or concerns held by all humans for all time \cite{macaskill2013infectiousness}. Value pluralism is widely adopted by queer theorists who highlight how formal value specifications typically exclude certain subpopulations in favor of others \cite{keyes2019counting}. For example, Kate Crawford has endorsed Mouffe’s (1999) concept of ``agonistic pluralism'' \cite{mouffe1999deliberative} as a design ideal for engineers \cite{crawford2016can}, while Hoffmann argues that abstract metrics of system fairness fail to address the hierarchical logic that produces advantaged and disadvantaged subjects and thereby disproportionately put safety harms on already vulnerable populations \cite{hoffmann2019fairness}. Mireille Hildebrandt has taken these perspectives to their logical extreme and advocates for ``agonistic machine learning'', suggesting that the human self should be treated as fundamentally incomputible \cite{hildebrandt2019privacy}. 

These conclusions have found support in the field of Computer Supported Cooperative Work (CSCW). Presenting them as a central challenge, Ackerman has described the inevitability of the ``social-technical gap'' of computer systems; the inherent divide between what we know we must support socially and what we can support technically \cite{ackerman2000intellectual}. This frames the central danger in terms of software engineers who neglect certain value hierarchies, either by failing to interrogate the context of historical data or external cost biases through design choices that moralize existing structural inequalities \cite{eubanks2018automating}. 
The call to value pluralism, as such, is not opposed to pragmatism in the form of external mechanisms that regulate how our diverse commitments may be reconciled \cite{james1896will}. Rather, as designers compromise the public interest through incomplete system specifications that create external costs for society, they have merely reframed the central problems of modern political theory \cite{dewey1954public} and inherited the hallmarks of structural inequality. 
The history of social technology, from the modern census to the invention of writing, is saturated with ways in which forms of human identity were problematically obfuscated or delimited rather than protected or left undetermined \cite{benjamin2019race}. 
This phenomenon underpins foundational concepts of twentieth century social theory \cite{krais1993gender} and deconstructionist critiques of Western philosophy as a ``metaphysics of presence'' \cite{heidegger1962being}.

On this view, any system design requires fundamental political choices about how values of relevant stakeholders, including those indirectly affected by the system, result in some value hierarchy that may have undesirable consequences for how the benefits and harms of a system are distributed across society. 
Correspondingly, the type of feedback most readily endorsed by ontic incomparabilists has been \textit{refusal}, i.e. the explicit rejection of a system specification as unsuitable. 
This has been expressed recently through comparisons of facial recognition systems with plutonium \cite{stark_facial_2019}, algorithmic classification with a new form of "Jim Code" \cite{benjamin_race_2020}, and refusal itself with the notion of feminist data practice \cite{garcia2020no}. 
However, a major open question is how or whether refusal itself can lead to the articulation of a more just and equitable society in the absence of alternative forms of feedback.

\subsection{Semantic indeterminism - declaring things fuzzy by nature}

Finally, \textit{semantic indeterminism} asserts that the extent to which we can determine the definition of a concept is the extent to which the members of a given community agree on that definition. 
Commonly associated with Wittgenstein \cite{wittgenstein1953philosophical}, this position emphasizes the rules of language-games as defining how we refer to the world and the specific boundaries of a given community’s concerns, social tastes, and modes of valuation. 
To again illustrate this via the Sorites paradox: Persians, Romans, and even distinct Greek city-states may use alternative definitions of ``heap'' and thus confidently draw different cutoff points without ontological disagreement. 
Semantic indeterminism does not argue for a radical version of social constructivism according to which any claim to describe reality is arbitrary or fictional, e.g. the notion that claims about the objective world are impossible. Rather, such claims simply cannot be interpreted outside the rules that particular language communities have adopted and refined over time.
 
Discursively, semantic indeterminism amounts to a belief in \textit{fuzziness}, the notion that the lines between our ways of talking about ``the world'' are blurry and spread unevenly between distinct language communities or modes of expertise. 
Lessig’s famous modalities of regulation (laws vs. norms vs. markets vs. architecture) are an example of the multiple ways of resolving this fuzziness \cite{lessig2009code}, as is ``fuzzy logic'' itself \cite{gerla2016comments}. 
In the context of AI Safety, an exemplary discourse has formed within the Fairness, Accountability and Transparency in Computing Systems (FAccT) literature. FAccT research has harvested a multitude of definitions and tools aiming to address safety risks by diagnosing and reducing biases across various subgroups defined along lines of race, gender or social class~\cite{narayanan_fat*_2018}. 

While scholars have pointed out the critical and mathematical shortcomings of abstract definitions for bias mitigation, these are still instrumented in practice as means to resolve fuzziness in particular application domains. For example, industry efforts have embraced bias tools to generate feedback for \textit{equitable outcomes} as a means to engender trust in a given system, while global efforts aim to codify algorithmic bias considerations into certifiable standards ``to address and eliminate issues of negative bias in the creation of [...] algorithms'' \cite{noauthor_ieee_2021}.
  
Still, the tension between eliminating bias and winning social trust reveals the inconsistent determinations of what safety means throughout the entire lifecycle, including which norms should guide design and use decisions. 
As some argue, ``it is important to acknowledge the semantic differences that `fairness' has inside and outside of ML communities, and the ways in which those differences have been used to abstract from and oversimplify social and historical contexts’’ \cite{rea2020survey}.
Scholars have also emphasized important semantic differences and connections between ``individual'' and ``social'' fairness that could help clarify and procedurally reshape the way formal fairness criteria are reconciled with policy objectives \cite{corbett2018measure,binns2018fairness}. However, incorporating these semantic differences would mean accommodating additional types of feedback, such as \textit{preference learning} to represent what people actually seem to want as well as \textit{refusal} to serve as a check on the system's tendency to occlude or suppress neglected values. Thus, semantic indeterminism does not resolve the normative indeterminacies raised by epistemicism and ontic incomparabilism, but rather defers them for either designers or stakeholders to deal with.
This is clearly exemplified in the EU's recent proposal for regulating AI systems in high-stakes domains, in which the need for ``harmonised standards'' is advocated, stating that ``[t]he precise technical solutions to achieve compliance with those requirements may be provided by standards or by other technical specifications or otherwise be developed in accordance with general engineering or scientific knowledge at the discretion of the provider of the AI system.''~\cite{noauthor_proposal_2021}

Instead, we propose that fuzziness is best understood as a \textit{sociotechnical} problem for AI development. This means that systems’ ``core interface consists of the relations between a nonhuman system and a human system'' \cite{trist1981evolution}, with various dimensions (e.g. users, citizens, operators, regulators), whose construction is hindered by limited knowledge, subject to error, of how key technical innovations bear on human contexts. Even carefully-designed formalisms that are sensitive to the implicit concerns of human agents are not guaranteed to learn the right preference structures in the right way without new forms of surveillance, control, and assigned roles for both humans and the systems themselves \cite{eckersley2018impossibility,agre1994surveillance}. Such system setups are limited in three ways: (1) they can never formalize everything, and require subsequent developers to organize around them; (2) they attempt to resolve (and thereby confuse) content and procedure from the get-go, rather than treat the sociotechnical development of AI systems as a dynamic problem; and (3) they are limited in addressing wider spectra of values across distinct peoples and cultures.

\begin{table}[h!]
    \centering
    \begin{tabular}{c|c|c}
        Type of Vagueness & Normative Standard & Mode of Feedback  \\
         \hline 
         Epistemicism & metanormativism & preference learning \\
         Ontic Incomparabilism & value pluralism & refusal\\
         Semantic Indeterminism & fuzziness & equitable outcomes
    \end{tabular}
    \label{tb:vague}
    \caption{Relationship between types of vagueness, the standard of normativity each assumes, and the types of feedback they each prioritize.}
    \label{tab:my_label}
\end{table}

\section{A Framework of Commitments for AI Development}
\label{sec:hcai}
As outlined in Section~\ref{sec:vagueness}, matching safety principles with technical development procedures is fraught with hard choices. There are inherent sources of vagueness about what safety means, how it is formalized, and how it is enacted in an AI system. 
As a result, indeterminacies are encountered through possible design interventions that are technically comparable but normatively incommensurable. 
If left unaddressed or underconsidered, these may lead to harms, reinforcement of structural inequalities, or unresolved conflict across different stakeholders. 
Section\ref{sec:vagueness} thus analyzed a broad spectrum of technical, governance and critical scholarship efforts to address the safety of AI systems, and how these fall in three canonical approaches to vagueness. 
For each lens, we determined the affordances and limitations of their associated cybernetic feedback modalities and the interventions that can be done with these to safeguard an AI system or improve the practices that design or govern it.

In this section we integrate these lessons, arguing that designers should address hard choices by incorporating appropriate types of stakeholder feedback into the development and governance of the system. 
We also build on those lessons by explicating the role of democratic dissent as a critical additional form of cybernetic feedback in AI system development and governance, as motivated in Section~\ref{sec:lexicon}.
Together, the facilitation of cybernetic feedback channels constitutes substantive commitments to the governance of the domain in which the system will operate. We thus delineate a set of commitments that would frame technical development as deliberative about the system’s normativity. This recasts the traditionally linear ``AI development pipeline'' process as dynamic and reflexive, comprising cybernetic design principles for AI governance.

\begin{figure}[!htb]
    \center{\includegraphics[width=12cm]
    {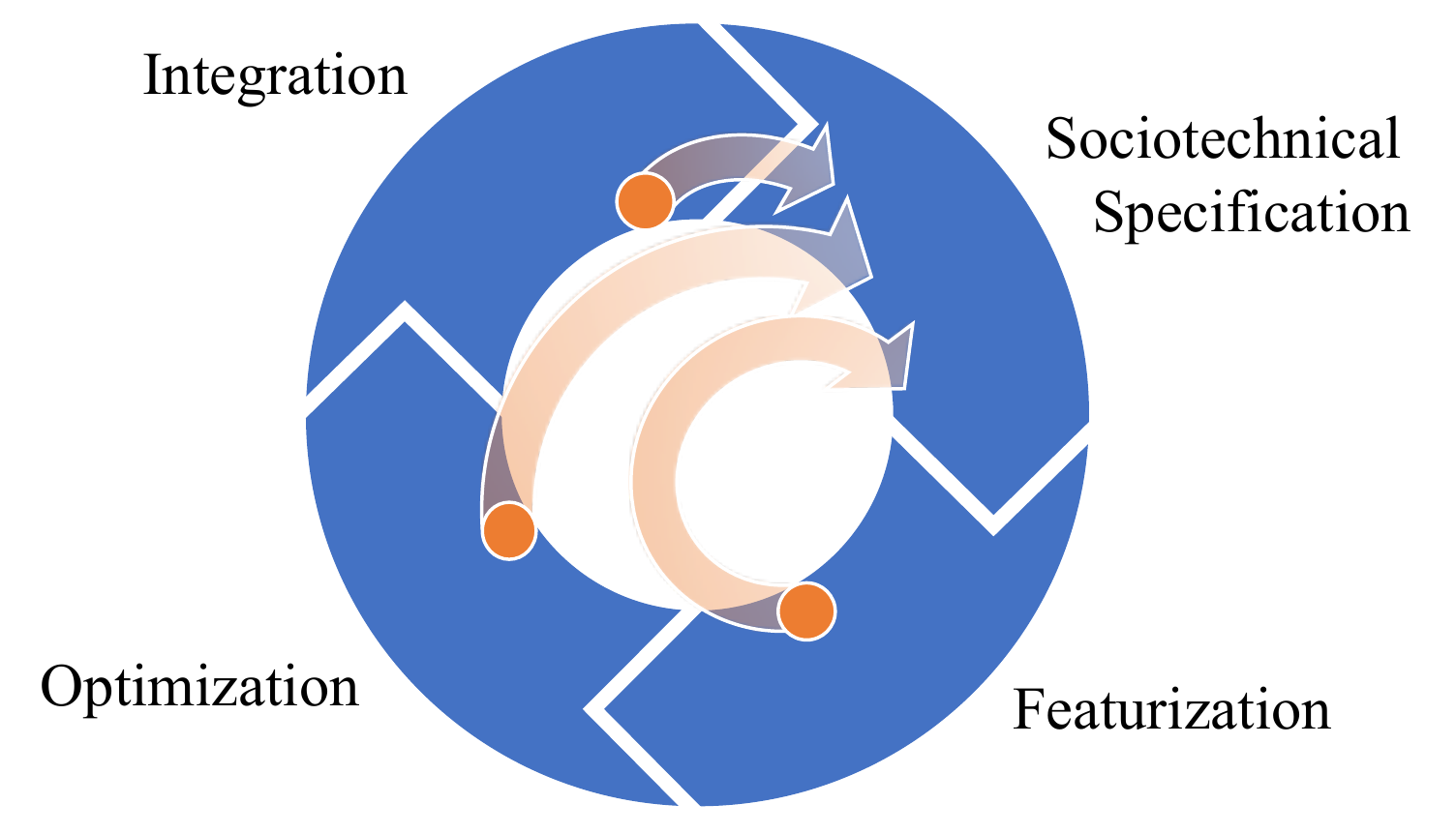}}
    \caption{\label{fig:my-label7} The cyclical practices in AI system development. Orange circles denote the occurrence of ``hard choices'', moments where normative indeterminacy arises, which require revisiting sociotechnical specification.}
\end{figure}

The resulting \textit{Hard Choices in AI (HCAI) Framework}, presented in Figure 1, contains four cybernetic practices: sociotechnical specification, featurization, optimization, and integration. These activities and corresponding commitments will be introduced and discussed in the following subsections. We stress that this framework is a conceptual depiction of how to deliberate critically and constructively about normative indeterminacy. The framework may however help to identify concrete design approaches that can put commitments in action. In many instances, regulatory measures may form either an existing source of constraints and requirements in the development process, or be informed by it. We do not advocate for particular law or policy interpretations, as these are just as contextual as design approaches, but see such translation work as a natural extension of this paper. Our framework naturally connects with and further concretizes the 'AI system lifecycle' as introduced in the OECD AI Principles~\cite{noauthor_recommendation_2021}.

\subsection{Sociotechnical Commitments}

Developers must diagnose situations of normative indeterminacy while remaining attentive to the fundamental limitations of technical logics to resolve them. This necessitates an ``alertness'' to all the factors responsible for the situation, including social, affective, corporeal, and political components \cite{amrute2019techno}. AI systems are not merely situated in some pre-existing sociotechnical environment. Rather, the development of the system itself creates novel situations that intervene on social life, reflected in the distinction between pre-existing, technical, and emergent bias \cite{friedman1996bias}. These require their own formal treatment \cite{dobbe2018broader}.
 
Furthermore, major stages of AI system development require feedback channels for stakeholders to assign appropriate meaning to possible specifications. In particular, we emphasize the need for dissent mechanisms to help surface parity of different design options and their related value hierarchies. As Anderson emphasizes, in contexts where a policy is set by a majority or powerful player, ``[s]uch dissent is needed not simply to keep the majority in check, but to ensure that decision-making is deliberative—undertaken in an experimental spirit—rather than simply imposed'' \cite{anderson2006epistemology}. These channels make AI development an opportunity for communities to reimagine their own moral boundaries.
 
Developers must also acquire practical reasoning to navigate across sociotechnical approaches to a problem and determine specifications accordingly. A specification that might make sense in one context may not make sense for another, either in terms of feature detection (e.g. facial vs. handwriting) or integration scale (municipal oversight vs. nationwide surveillance). Developers must recognize the differences between these and internalize standards that guide the indeterminate application of abstract principles to the concrete needs and demands of the situation, in a manner responsive to stakeholder feedback. These comprise distinct forms of judgment: formulating the problem, evaluating system criteria, and articulating the performance thresholds that the system must meet in order to be safe. We agree with Philip Agre that this engagement requires ``reflexive inquiry [that] places all of its concepts and methods at risk [...] not as a threat to rationality but as a promise of a better way of doing things'' \cite{agre_toward_1997}.
 
At distinct moments of formal specification, we ask: (1) at what development stages and associated cybernetic practice might indeterminacies manifest and what forms may parity take? (2) in what concrete ways are feedback mechanisms/interventions needed to address these issues? (3) what form does the associated canonical dilemma take? (4) what forms of judgment are needed to interpret stakeholder feedback and effectively manage the indeterminacies and dilemmas that the system generates? These points are presented in Table 2.
\begin{table}[h!]
    \centering
    \begin{tabular}{c|c|c|c}
        Cybernetic Practice & Intervention & Dilemma & Sociotechnical Judgment  \\
         \hline 
         Sociotechnical Specification & integral &  inclusion vs. resolution & solidarity \\
         Featurization & epistemic & underfeaturized vs. misfeaturized & context discernment\\
         Optimization & semantic & verification vs. validation & stewardship\\
         Integration & ontic & exit vs. voice & public accountability
    \end{tabular}
    \label{tb:vague}
    \caption{Relationship between cybernetic practices, normative interventions, hard choice moments requiring feedback, and forms of sociotechnical judgment needed to interpret feedback.}
    \label{tab:my_label}
\end{table}

\subsection{Sociotechnical Specification (engaging the ``stakes'' and forms of agency)}

The HCAI Framework does not identify a clear start of AI development, but it does require the initial determination of how the problem is to be formulated and tackled, mechanisms for improving this determination through feedback and dissent, and what stakeholders are already implicated or should be involved in problem formulation. 
Moreover, not all normative dimensions can be foreseen upfront, as hard choices may surface in subsequent development considerations. 
Aware of these historical, critical, and empirical complexities, we center the need for \textit{sociotechnical specification}, i.e. the process of facilitating the different interests relevant in understanding a situation that may benefit from a technological intervention. Developers must clarify what the system is actually for--whose agency it is intended to serve, who will administer it, and what mechanisms are necessary to ensure its operational integrity. The sociotechnical specification facilitates \emph{integral} interventions to determine and resolve what safety means (semantic), how it is formalized (epistemic), and how it is enacted in a system (ontic). This facilitation cannot fall exclusively on the plate of designers or developers. 

To appropriately surface parity throughout sociotechnical specification, the following challenges must be taken up: (1) negotiate a program of requirements and conditions on both process and outcomes; (2) determine roles and responsibilities across stakeholders; (3) agree on ethics and modes of inquiry, deliberation, and decision-making. In sociotechnical specification, one needs to understand the context of integration. This includes the positions of different stakeholders with their reasoning and how these relate to each other. It requires an understanding or anticipation of the impacts on social behavior, broader societal implications, and how different solutions would sit within existing legal frameworks. This yields the following dilemma:

\textit{Inclusion}: What stakeholders are directly involved or indirectly affected by issues and solution directions considered? How is power and agency assigned along the process of development and integration? How are the boundaries of the AI system and its implications determined? 

\textit{Resolution}: What deliverables or outcomes are expected or envisioned for the project? What variables and criteria are needed to measure these outcomes? What ethical principles and decision-making process is needed to achieve resolution across different stakeholders? What conditions will allow both supportive and dissenting groups to express their concerns and contribute meaningfully to the development and integration of a resulting system?

The key hard choice for a successful AI system is to include sufficient perspectives and distribute decision-making power broadly enough in development to cultivate trust and reach a legitimate consensus, while resolving the situation in a set of requirements and a process with roles and responsibilities that are feasible. While we propose these diagnostic and procedural questions for AI system applications broadly (and prospectively for more computationally intensive systems in the future), here we focus our attention on contexts that are safety-critical by nature or play an important public infrastructural role. This includes systems that integrate on a global scale, interacting with a wide spectrum of local and cultural contexts.

\textit{Solidarity} is necessary to resolve this hard choice by specifying warranted interventions for the system’s subsequent development. The criterion for these interventions as warranted is twofold. First, indeterminacies that would necessarily prevent the system’s successful operation must be resolved in advance. Second, indeterminacies that do not threaten successful operation must be deferred for stakeholders to evaluate and interpret according to their own involvement and concerns. In this way, interventions will align abstract development commitments with specific possible design decisions, given the particularities of the situation and the most urgent needs of relevant stakeholders. 
Indeed, the three subspecies of hard choices described below do not comprise a linear, abstract checklist so much as forms of situational alertness to the possibility of parity throughout the iterative development process. Ideally, the initial problematization stage identifies all the strategies and modes of inquiry necessary to track and resolve indeterminacies. This includes an appropriate assignment of roles and responsibilities across all stakeholders.

Solidarity should not be understood as conflating the interests of designers and stakeholders. Rather it motivates the former to create channels for stakeholders to actively determine, rather than passively accept, the system specification \cite{unger1983critical}. Here we endorse Irani et al's vision for postcolonial computing, which ``acknowledge[s] stakeholders as active participants and partners rather than passive repositories of ‘lore’ to be mined'' \cite{irani2010postcolonial}.

\subsection{Featurization (epistemic uncertainty)}

AI systems generally represent a predictive, causal or rule-based model, or a combination thereof, that is then optimized and integrated in the decision making capabilities of some human agent or automated control system. As such, it has to answer the question ‘what information it needs to ``know'' to make adequate decisions or predictions about its subjects and notions of safety?’. As the model represents an abstraction of the phenomenon about which it makes predictions, the chosen model parameterization and the data used to determine parameter values delimit the possible features and value hierarchies that may be encoded. If not anticipated and accounted for, this may deny stakeholders the opportunity to evaluate design alternatives and force potentially harmful and unsafe hard choices. In this way, featurization is an \textit{epistemic} intervention on the indeterminacies that may be present or latent in the context that precedes or follows system operation.
 
To surface the parity at stake in featurization, the following challenges must be taken up: (1) make explicit and negotiate what can and cannot be modeled and inferred, crystallized in the underfeaturized/misfeaturized hard choice; (2) engage stakeholders to challenge and inspire modeling assumptions to ensure application aligns with contextual expectations; (3) validate the design with stakeholders to anticipate possible value conflicts that can arise due to the gap between model and world and plurality of values during deployment, preparing to revisit the modeling tools and methodology. Featurization specifies the computational powers of the system: how the limits of what it can model determine its assumptions about people and the broader environment, and what kinds of objects or classes are recognizable to it. At a minimum, stakeholders must resolve the following dilemma: 

\textit{Underfeaturized}: What possible input variables or model parameterizations do we choose not to include? What features will the model not be able to learn that may in fact be open to normative deliberation?

\textit{Misfeaturized}: What environmental features or actions do we choose to parameterize, and with what complexity? What forms of dissent will be foreclosed by elements of computation, and for whom would this matter?

The danger lies in failing to adopt model parameters that are both computationally tractable and normatively defensible. Given finite time and material resources as well as the vested interests of specific stakeholders, this may err towards under- or mis-specification in ways that developers cannot perfectly anticipate. The spirit of the hard choice is crystallized differently in distinct algorithmic learning procedures. For example, the division between model-based and model-free reinforcement learning essentially bears on what kind of control system is being designed and, respectively, whether this specification establishes a permissible space in which a given problem can be formulated and represented causally or merely defines permissible predictive signals (e.g. rewards, elements, qualities) within the environment. 
At least some corresponding domain features may be made computationally tractable and suited to optimization despite being experienced by stakeholders as incommensurable. Or some features may be technically obfuscated despite their mutual comparability and integrity in lived experience. 
An often returning example of this dilemma is the need to interpret or explain the decision-logic of an AI model.
While deep learning models may offer a higher performance, this need may lead to opting for a lower complexity model that has more potential for forms of accountability.
 
The model must be capacious enough to represent the nature of the environment in a way that safeguards stakeholders' interests. But its training must also be constrained enough to be tractable, guarantee performance \cite{achiam2017constrained}, and preserve privacy boundaries. Imposing modeling constraints necessarily creates technical bias, which may take away space for stakeholders to express or protect their own specific values in terms of the phenomena permitted or excluded by the model's system boundaries \cite{dobbe2018broader}. There is already some technical work acknowledging this as a formal dilemma with no optimal solution in the context of reinforcement learning \cite{choudhury2019utility, yu2019meta}. But the deeper sociotechnical point is that the criterion for these constraints, which entail a choice of the moment at which a model must remain technically ignorant or intentionally suboptimal, must be specified in terms of a commitment to the self-determination of stakeholders.
 
 
Featurization requires \textit{context discernment}, the disqualification of specific features and modeling choices that, while technically proficient, are judged to be sociotechnically inappropriate within the problem space at hand. Here we draw from \cite{dreyfus2011all}: ``The task of the craftsman is not to \textit{generate} the meaning, but rather to \textit{cultivate} in himself the skill for \textit{discerning} the meanings that are \textit{already there}.'' Featurization is about anticipating how the model would interact with the context of deployment, how else it could be (mis)used, what bias issues may arise during training, how to protect vulnerable affected groups, and how learned objective functions may generate externalities. In the event no consensus is reached and dissent persists, the option of not designing the system should be preserved \cite{baumer2011implication}.

\subsection{Optimization (semantic indeterminacy)}

The parameters of the system's internal model must be further determined by performing some form of optimization. This determines the input-output behavior of the model and how it will interact with human agents and other systems. Optimization extends across the design stage (e.g. training an algorithm) and implementation (e.g. finetuning parameters) and answers the question ‘what criteria and specifications are considered to measure and determine whether a system is safe to integrate?’. Depending on the chosen representation, such optimization can either be performed mathematically, done manually through the use of heuristics and tuning, or some combination thereof. For mathematical optimization, the recruitment of historical and experimental data is needed to either (a) infer causal model parameters (e.g. for system identification, an inference practice common in control engineering \cite{guo2018survey}), (b) infer parameters of noncausal representations, or (c) iteratively adjust parameters based on feedback (as in reinforcement learning). The objectives and constraints and the choice of parameters constitute a \textit{semantic} intervention on how the identification of specific objects relates to the forms of meaning inherited by and active in the behavior of stakeholders themselves.
 
Therefore the following challenges must be taken up: (1) assess the extent and limitations with which the optimization criteria and procedure can translate and respect specifications, crystallized in the validation/verification tradeoff; (2) codify a validation procedure for empirical criteria that conforms to stakeholders’ specific concerns, addressing specifications not covered through mathematical optimization; (3) adjudicate and modify verification and validation strategies over time as indeterminacies of featurization and integration continue to be highlighted. To declare a system safe it must go through a process of verifying and validating its functionality, both of itself as an artifact as well as integrated in the context of deployment. This is done with the help of engineers and domain experts who interface between the problem the system is meant to solve and the workings of the system itself. Here, the minimum requirements for safe outcomes are impartial assessments of the following questions/dilemma:
 
\textit{Verification}: Does the system meet its specifications (was the right system built)? Are the needs of prospective users being met? Is the system able to predict or determine what it was meant to?
 
\textit{Validation}: How does the system perform in its empirical context (was the system built right)? Does the system behave safely and reliably in interaction with other systems, human operators and other human agents? Is there risk of strategic behavior, manipulation, or unwarranted surveillance? Are there emergent biases, overlooked specifications, or other externalities? 
 
This hard choice poses several concrete challenges for development. First, systems that are mostly optimized in a design or laboratory environment fall inherently short as their data cannot fully capture the context of integration. In the development of safety-critical systems, this design issue is acknowledged by the need to minimize any remaining errors in practice (through feedback control \cite{aastrom2010feedback}) and putting in place failsafe procedures and organizational measures as well as promoting a safety culture. Second, accounting for interactions with other systems and human agents is not to be taken lightly and is heavily undervalued in current AI literature \cite{parasuraman1997humans}. For example, the overspecification of environments through simulation (as is now popular in the development of autonomous vehicles) may backfire if the optimization scheme overfits the model for features or elements that are not reflective of the context of integration. Third, a lack of validation and safeguarding systems in practice can result in disparate impacts \cite{barocas2016big} and failures. This is especially pertinent for underrepresented (and undersampled) groups that are often not properly represented on AI design teams \cite{west2019discriminating}. For systems that are ``optimized in the wild'' with reinforcement and online learning techniques, these considerations are even more acute, although recent efforts have proposed hybrid methods that can switch from learning to safety-control to prevent disasters \cite{fisac2018general}. This technical point, which mirrors the well known bias-variance tradeoff, becomes \textit{sociotechnical} at the moment when the choice of optimization procedure is interpreted from the standpoint of jurisprudence applicable to the domain.

 
The cultivation of \textit{stewardship} is needed to reconcile the technical problematics of value alignment with optimization procedures capable of providing qualitative assurances to the particular sociotechnical stakes of the domain, whether physical, psychological, social, or environmental. System engineers must internalize an understanding of how the finitude of their teams’ tools and procedures bears on the urgency felt by stakeholders towards objects of sociotechnical concern, compelling attention to how sparse team resources should be allocated and complemented, rather than to abstract notions of accuracy or efficiency. Only in this way can under- or mis-featurization risks be managed and mitigated without perverting intended stakeholders’ semantic and moral commitments. The team must decide: what internal verification strategies might we need in order to safeguard the validations already endorsed by legal inquiry? Here ``quality management'' must be elevated to the contestation and adjudication of how (possibly pluralist) values are operationalized without compromising parity.

\subsection{Integration (ontic incomparabilism)}

Finally, as AI systems are rapidly introduced into new contexts, new forms of harm emerge that do not always meet standard definitions. In addition, the diversity of stakeholder expectations, as well as of environmental contexts, may challenge specifying safety for systems that are deployed across different jurisdictions. At a minimum, those developing and/or managing the system must specify mechanisms to identify, contest, and mitigate safety risks across all affected communities, as well as who is responsible for mitigating harms in the event of accidents. This can be done via general rules and use cases of safety hazards that identify terms of consent, ensure interpretive understanding without coercion, and outline failsafe mechanisms and responsibilities. Hence, such conditions should spell out both the technical mechanisms as well as the processes, organizational measures, responsibilities, and cultural norms required to prevent failures and minimize damage and harm in the event of accidents. Here we appropriate tradeoffs already identified by social theorists regarding the moral authority and political powers of social institutions \cite{flew2009citizen}. This dimension serves as a decisive \textit{ontic intervention} of what kind(s) of agency stakeholders possess as far as the system is concerned.
 
To safeguard parity at integration, the following challenges have to be taken up: (1) assess what kind(s) of agency all affected stakeholders have if the system fails, crystallized in the exit/voice hard choice; (2) establish open feedback channels by which stakeholders express their values and concerns on their terms; (3) justify these channels as trustworthy through regular public communication and updates to the design and/or governance of the system. Resolving these challenges requires representative input and mitigation of issues for the following dilemma:
 
\textit{Exit}: Are stakeholders able to withdraw fully from using or participating in the system? Is there any risk in doing so? Are there competing products, platforms or systems they can use? Have assurances been given about user data, optimization, and certification after someone withdraws?
 
\textit{Voice}: Can stakeholders articulate proposals in a way that makes certain concerns a matter of public interest? Are clear proposal channels provided for stakeholders, and are they given the opportunity to contribute regularly? Are the proposals highlighted frequently considered and tested, e.g. through system safety? Are stakeholders kept informed and regularly updated?
 
To the extent that proposed value hierarchies remain indeterminate beyond featurization and optimization, sociotechnical integration challenges systems to handle the multiple objectives, values, and priorities of diverse stakeholders. At stake here are the unexpressed moral relationships of subpopulations not originally considered part of the potential user base, who must bear the ``cost function'' of specification, as well as other forms of agency (animal, environmental, cybernetic) alien to yet implicated in system specification and creation. At a minimum, system administrators must acknowledge that users will interpret the system agreement both as economic (acting as a \textit{consumer}) as well as political (acting as a \textit{citizen}). The developments on social media in recent years have taught us that these roles cannot be seen as mutually exclusive. The increasing dependence of the public on these platforms and their AI systems strengthens the need for voice (as exit options have become increasingly difficult or unlikely). 

 
Administrators must cultivate \textit{public accountability} to deal with these challenges, ensuring both Voice and Exit remain possible for stakeholders such that some criterion of trustworthiness is maintained. That is, anyone can leave the service contract if they want, but enough people choose to remain because they believe in their ability to express concerns as needed. Trustworthiness lies in supporting stakeholders' belief in their ability to exert different kinds of agency as they see fit, either within the system (by dissenting to its current mode of operation) or outside it (by choosing it through active use). This sociotechnical balance must hold regardless of the specific commitment being made. For example, service providers may specify some channel by which vulnerable groups can opt out of a publicly-operated facial recognition system (preserving Exit), or supply private contractors with a default user agreement that must be relayed to anyone whose data will be used by the system (preserving Voice). Either way, administrators must ensure they treat people both as respected consumers (a customer, client, or operator treated more or less as a black box) as well as citizens (a subject with guaranteed rights, among them the right to dissent to relevant forms of political power) in the context of the terms for system integration. Failure to have meaningful exit or voice can motivate collective action to reshape power relationships \cite{hirschman1970exit}, a phenomenon that has recently manifested when pushing back against harmful AI systems \cite{crawford_ai_2019}.

\section{Implications and Discussion}

HCAI serves as a systematic depiction of the normative risks and sociotechnical gaps at stake in any AI system. But how should developers respond when examining particular proposed or existing systems? Here we present the normative implications of HCAI in terms of practical recommendations that go beyond existing governance and performance standards. We identify opportunities for policymakers, AI designers, and STS scholars to learn from each others' insights and adopt a cohesive approach to development decisions.

\subsubsection*{Expand the boundary of analysis to include relevant sociotechnics - systems, organizations and institutions}
Engineering and computer science disciplines have long tradition of working with ``control volumes'', which are mathematical abstractions employed to render problems and their solutions in terms of technical terms~\cite{li_will_2007}. In doing so, they allow a designer to decontextualize, depoliticize and ignore the history of a problem~\cite{kadir_engineering_2021}. 
While often done in a more controlled context, the sociotechnical complexity and normative stakes of AI systems engaging in sensitive social and safety-critical domains requires a more comprehensive lens.
An algorithm or AI system alone cannot engage with its inherent normativity.
In contrast, studies in systems safety have shown that safety is inherently an \emph{emergent property} that ``arises from the interactions among the system components.''~\cite{leveson_engineering_2012} This requires a system perspective that includes the human agents interacting with a technology~\cite{green_principles_2019}, as well as how it is situated with respect to organizational processes~\cite{von_krogh_artificial_2018} and cultural and institutional norms~\cite{gasser_role_2020}.
Such a systems lens also provides a more comprehensive starting point for controlling for safety, which is done by ``imposing constraints on the behavior of and interactions among the components'' of a system~\cite{leveson_engineering_2012}. This lens also explains how vulnerabilities of AI systems originate from across these components and system interactions, which corroborates insights from computer security that systems cannot be secured by addressing technical/mathematical vulnerabilities alone~\cite{crawford_ai_2019,carlini_evaluating_2019}.
Lastly, a broader systems lens will be vital in understanding to what extent intended standards for AI systems~\cite{noauthor_proposal_2021} can lean on general principles versus contextual needs and stakes specific to the domain of application. 
AI developers can lean on a long history in systems engineering of analyzing, modeling and designing sociotechnical systems, which should go hand-in-hand with a multi-actor approach~\cite{de_bruijn_system_2009}.

\subsubsection*{Confront the choices and assumptions behind the AI system}
Rather than addressing the limitations of a formalization itself, an honest encounter with normative indeterminacy deserves an account of the normative assumptions behind it and their implications.
Recently, various scholars have advocated about the dangers of abstraction in AI systems~\cite{selbst_fairness_2019}, and pointed to the dangers of how imposing such abstractions can reify inequities resulting from institutional racism~\cite{benjamin_race_2019} and harm marginalized communities~\cite{bender_dangers_2021}. Put bluntly, the choice of capturing everything in terms of an AI model and objective function is political. Cybernetic practices should make explicit where the boundary for acceptable formalization lies and what forms of feedback and evaluation are needed to safeguard their integration.
Apart from their role in formalizing, modeling choices come with their own externalities. An inverse reinforcement learning procedure inherently requires the observation of detailed human behavior which might violate privacy norms~\cite{raji_concrete_2020}. And deep learning architectures are synonymous with extensive data gathering, which challenges privacy as well as environmental norms~\cite{dobbe_ai_2019}.

\subsubsection*{Orient development and governance towards a multi-actor approach with ``problem and solution spaces''}

As De Bruijn and Herder argue, in addition to a more techno-rational systems lens, one needs to take into account the effects of different intentions of actors involved in or affected by a system~\cite{de_bruijn_system_2009}. 
As we saw in our vagueness analysis, actors may have different lived realities, languages or epistemic perspectives and, as a result, conflicting interests or incommensurable demands. 
Generally speaking, the actor perspective acknowledges and conceptualizes the dependencies between actors, sometimes captured in an ``issue network''~\cite{borzel_organizing_1998}, and develops the ``rules of the game'' or governance mechanisms needed to satisfy all actors and manage the system adequately.
While it is obvious that the system and actor perspective should at least happen in parallel and be in conversation, opinions vary on how integrated they should be, which in itself is a matter of normative indeterminacy. 
In this paper we argue that diagnosing and grappling with normative indeterminacy (or providing design space for parity) requires cybernetic feedback, which we specify both at the system (dynamical feedback) and at the actor level (feedback to renegotiate what abstractions and procedures are necessary to safeguard a system). 
The iterative nature of dealing with emergent hard choices in AI system development, requires an iterative approach that also revisits the stakes and consensus reached among actors. 
As such, AI safety can only emerge through a \emph{reciprocal relationship between system development and governance}. 

Complex multi-actor problems are often called \emph{wicked problems}, especially when they are subject to normative indeterminacy: ``Wicked problems have incomplete, contradictory, and changing requirements, and solutions to them are often difficult to recognize as such because of complex dependencies.''~\cite{de_bruijn_system_2009} Put differently, ``they rely upon elusive political judgment for resolution. (Not ``solution.'' Social problems are never solved. At best they are only re-solved--over and over again.) [..] The formulation of a wicked problem is the problem!''~\cite{rittel_dilemmas_1973}
At a minimum, a safety-critical context requires an honest account of the \emph{problem and solution spaces}, which elaborate the different perspectives on the problem and its solution by various actors, as a basis for trying to reach broad consensus~\cite{de_haan_solving_2015}.

\subsubsection*{Acknowledge the connections between specification and political interests}
Because the value hierarchy specified for and designed into a system will determine the space of actions available to it (as well as those that the system forecloses), it is crucial to acknowledge and account for the power and elevated status of design work~\cite{irani2016stories}. This means recognizing developers' tendencies to prioritize certain actors and networks over others. Haraway~\cite{haraway1988situated}, Harding~\cite{harding1986science}, and other critical scholars would argue that we cannot escape having some agenda: researchers are themselves situated in the social world they study. As such, technology development is inherently political and requires forms of accountability~\cite{wagner_accountability_2020}. 
Pioneers in participatory design argue that conflicts should be expected and that ``[i]t’s not the IT designer’s job to cover up or try to solve political conflicts that surface [...] it is their job to develop different design visions and assess their consequences for the affected parties.''~\cite{bodker2009participatory} However, there are recent concerns that design methods using participation as a form of accountability are increasingly co-opted and stripped of their essence~\cite{bodker_participatory_2018,bannon_reimagining_2018}.

However, reducing political reflection to the role of the ``developer'' is too narrow to adequately capture the implications for specification.
Just like other actors, developers are embedded in a network and subject to power differentials themselves. 
Understanding how broader hierarchies of power both promote and constrain certain problem formulations is necessary to determine viable strategies for promoting system safeguards.
Today, much AI research and development, system implementation and management, as well as computational and software infrastructure is in the hands of a small number of technology companies. 
As G{\"u}rses and Van Hoboken argue, the move of tech companies to offer software engineering tools and data provision in service libraries and APIs has made the development of ``values by design'' an elusive task, and enabled new economic feedback loops that, when implemented at scale, drive new forms of inequality across social groups~\cite{gurses_privacy_2017,kostova_privacy_2020}. 
We believe that real success in safeguarding high-stakes systems will require democratic structures and procedures to meaningfully engage and impose forms of oversight and dissent that have the teeth to adequately respond to emergent safety hazards, especially for AI systems developed and deployed by the private sector and state actors.


\section{Conclusion}

Our framework is strongly influenced by the classic work of Philip Agre, which aimed to have AI practitioners and designers build better AI systems by requiring ``a split identity - one foot planted in the craft work of design and the other foot planted in the reflexive work of critique''. While we embrace the spirit of Agre’s work, we also believe that the critical applications of today’s AI systems require a new lens that can see beyond technical practices, and reframes the inherently interdisciplinary practice of AI development as critical in its own right. Apart from reflexivity, such a critical practice includes the forms of feedback that the domain of application asks for. The technical work done by AI practitioners plays a necessary but not sufficient part in development. It must be compensated by efforts to facilitate stakeholders’ ability to be ``full and active participants,'' while ``the tools and techniques for doing this are dependent on the situations within the workplace...steer[ing] toward understanding different, pluralistic perspectives of how we think and act'' \cite{greenbaum_design_1992}. As such, we prioritize and label the centering of stakeholder safety concerns and hard choices to guide and inform AI development as \textit{cybernetic practices}. We view this paper as a preliminary for what forms these practices might take in particular development domains, and will pursue this effort in future work.

Our lodestar in this project is the intuition that clarifying the sociotechnical foundations of safety requirements will lay the groundwork for developers to take part in distinct dissent channels proactively, before the risks posed by AI systems become technically or politically insurmountable. We anticipate that cybernetic practices will need to be included within the training of engineers, data scientists, and designers as qualifications for the operation and management of advanced AI systems in the wild. Ultimately, the public itself must be educated about the assumptions, abilities, and limitations of these systems so that informed dissent will be made desirable and attainable as systems are being deployed. Deliberation is thus the goal of AI Safety, not just the procedure by which it is ensured. We endorse this approach due to the computationally underdetermined, semantically indeterminate, and politically obfuscated value hierarchies that will continue to define diverse social orders both now and in the future. Democratic dissent is necessary for such systems to safeguard the possibility of parity throughout their development and allow users to define the contours of their own values. To paraphrase Reinhold Niebuhr \cite{niebuhr1986essential}, AI's capacity for specification makes hard choices possible, but its inclination to misspecification makes them necessary.

\section*{Acknowledgements}  We wish to thank Michael Dennis, Joan Greenbaum, Iason Gabriel, Rashida Richardson, Elizabeth Kaziunas, David Krueger, \'{I}\~{n}igo Mart\'{i}nez de Rituerto de Troya, and Seda Gürses for their constructive feedback on earlier versions of this paper. T.K. Gilbert is funded by the Center for Human-Compatible AI as well as a Newcombe Fellowship.


\bibliographystyle{apalike}
\bibliography{mybibfile,references_hc}

\end{document}